# Efficient Clear Air Turbulence Avoidance Algorithms using IoT for Commercial Aviation

Amlan Chatterjee, Hugo Flores, Bin Tang
Department of Computer Science
California State University, Dominguez Hills
Carson, CA USA
{achatterjee,btang}@csudh.edu
hflores27@toromail.csudh.edu

Ashish Mani[+], Khondker S. Hasan[*]
Dept. of EEE, Amity University, India[+]
College of Science & Engineering, UHCL
Houston, TX, USA[*]
amani @amity.edu[+]
HasanK@uhcl.edu[*]

*Abstract*— With the growth of commercial aviation over the last few decades there have been many applications designed to improve the efficiency of flight operations as well as safety and security. A number of these applications are based on the gathered data from flights; the data is usually acquired from the various sensors available on the aircraft. There are numerous sensors among the electrical and electronics devices on an aircraft, most of which are essential for the proper functioning of the same. With the sensors being operational throughout the time of movement of the aircraft, a large amount of data is collected during each flight. Normally, most of the gathered data are stored on a storage device on the aircraft, and are analyzed and studied later off-site for research purposes focusing on improving airline operation and efficiently maintaining the same. In certain cases, when there is data transfer during the flight, it is between the aircraft and an air-traffic-control (ATC) tower, which serves as the base station. The aircraft equipped with all these sensors, which can gather and exchange data, form a framework of Internet of things (IoT). Detecting and avoiding any form of turbulence for an aircraft is vital; it adds to the safety of both passengers and aircraft while reducing the operating cost of the airline. Therefore, in this paper, we study techniques to detect and avoid Clear Air Turbulence (CAT), which is a specific type of turbulence, based on the IoT framework of aircraft. We propose algorithms that consider both direct and indirect communication between aircraft within a specific region. Using simulation results, we show that our proposed techniques of direct communication using the IoT framework is faster than conventional techniques involving radio communication via both single ATC tower and multiple ATC towers.

## I. Introduction

Air travel has always been the preferred mode of transportation with respect to time and safety considerations. With commercial aviation increasing the number of flights each year, the challenges of maintaining the stringent requirements of operation are studied now more than ever. In the United States, currently there are about 5,000 flights airborne at any given time. There are more than 43,000 flights handled daily on an average by the Federal Aviation Agency (FAA), and the total number of flights handled by the FAA in 2016 was more than 16 million, operating from more than 19,000 US airports [12].

In the era of IoT, frameworks have been developed and already in use in major airports to help with luggage handling, tracking, enhance passenger experience etc. [11][16]. Flights also gather and store huge amounts of data along the flight paths during the trips. These data sets are analyzed later for insights into improving the efficiency of flight operations [4][5][6]. However, data gathered during flights might also be useful in certain cases if analyzed real-time rather than later and shared among other airborne aircraft within the region. There are options of communicating with the air traffic control towers, that act as the base stations, to send over such gathered data during the flight. However, to improve flight operations, it is imperative that flights are able to communicate directly with each other. In addition, if all flights communicate with the base station, or use the base station as an intermediary between them, it will become a bottleneck for the system; there would be increased workload for the ATC towers as well as increased latency for the data transmission. Also, to reduce the effect of single point of failure, the data should be offloaded as it gets generated. Compressing the transferred data can be one technique to reduce the network bandwidth, but the overhead adds to the latency [3][9].

In this paper, we propose a framework for IoT with regards to aircraft and associated sensor & communication devices. We specifically focus on detection and avoidance of turbulence in commercial aviation. Generally, there are four types of turbulence that an aircraft might encounter during flight: thunderstorms, mountain wave turbulence, wake vortex and clear air turbulence. The topic of focus for this paper is clear air turbulence (CAT) avoidance. We introduce algorithms to detect CAT that are based on both direct and indirect communication among aircraft. The proposed techniques show that, in general direct communication between aircraft using IoT framework is the most efficient technique of avoiding CAT once it is detected. The same principles can also be applied to detect and avoid the other types of turbulence as well.

The outline of our paper is as follows. In Section II, we present information on previous work related to different techniques for CAT detection and avoidance in commercial aviation. Section III provides the basic information regarding airspace sharing and the impact of turbulence on aircraft within a region. In Section IV, we introduce the different scenarios for clear air turbulence detection algorithms. The techniques to detect and avoid CAT using IoT devices abroad aircraft are introduced in Section V. Results of the implementation of the introduced algorithms for CAT detection and avoidance is presented in Section VI. Conclusion and future work is discussed in Section VII.

## II. RELATED WORK

Detecting air turbulence is a well-studied phenomenon, and as a part of it clear air turbulence has been investigated as well. There have been previous research studies that have discussed air turbulence with respect to aviation [15] and specifically the different aspects of CAT.

Using radar to track aircraft and detect turbulence has been proposed before. Hence, techniques to optimize radar detection of CAT also exist [1]; however, these methods do not work in radar shadow zones and clutter zones.

There have been enhancements proposed to radar resulting in the usage of advanced technology such as LIDAR (Light Detection and Ranging) systems, but they are subject to errors, specifically due to horizontal winds at high altitudes [13][14].

Methods to predict CATs with certain probability also exist [10]. However, since commercial aviation incurs heavy damage if passengers or crew are injured, minimizing such incidents using definitive techniques are required.

Related to aviation there are applications that perform analysis on airline data using IoT devices. Automatic dependent surveillance-broadcast (ADS-B) data exchanges between receivers and equipped aircraft have been analyzed to be used in applications including airspace and traffic monitoring [7].

To the best of our knowledge, this is the first work that proposes using IoT based techniques to detect CAT. There are other distributed sensor and crowdsourcing applications in different domains that use similar basic principle. For example, there are ways to detect potholes on roads using sensor based applications [8]. Although this is based on detecting abnormal conditions using sensors on existing devices [2], the challenges involved in commercial aviation are fundamentally different. In this paper, we consider scenarios of detecting CAT using both direct and indirect message transfer between nearby aircraft.

## III. AIRSPACE SHARING AND IMPACT OF TURBULENCE

Commercial aviation has advanced over the last few decades, and there has been a huge increase in the number of flights. Thus, the demand to share airspace has been greater than ever. There are three different types of movement for aircraft: vertical, horizontal and lateral. To facilitate the sharing of airspace, there are guidelines specifying the minimum separation of aircraft in the direction of each of the three movements. The vertical separation is set at 1000 feet, but often 2000 feet is used as an added precautionary measure. The lateral separation is 50 miles, and the horizontal separation i.e., the space between two consecutive flights on the same path is usually kept at a minimum of 10 minutes of travel distance.

With flights taking different altitudes during travel, there can be different tracks on the same flight path separated vertically. To facilitate safe sharing of airspace between flights, it is imperative to track the exact location of flights at all times and be aware of all other flights within a specific distance. Flights originating and terminating over land routes can maintain communication with ATC over radio and their precise locations can be tracked via radar. Using the location information, messages can be sent to flights about sudden changes in weather conditions. However, for flights taking oceanic routes from one continent to another often are outside of any radar coverage and in limited communication zones. Therefore, in such cases, communication using conventional radio technology is not reliable. Now, at a given time, there can potentially be a large number of aircraft within a region of similar weather conditions. Therefore, if certain aircraft detect turbulence, specifically CAT, then the probability of other aircraft in the same region experiencing turbulence is reasonably high. Hence, it is significant in the domain of commercial aviation to be able to detect and essentially avoid any such turbulent routes if possible. So, in this paper, we consider the different scenarios to detect CAT and methods to take preventive action. Specifically, if CAT is detected by an aircraft, then a message indicating the location must be sent to all other aircraft within the same airspace to minimize the potentially hazardous effects of turbulence.

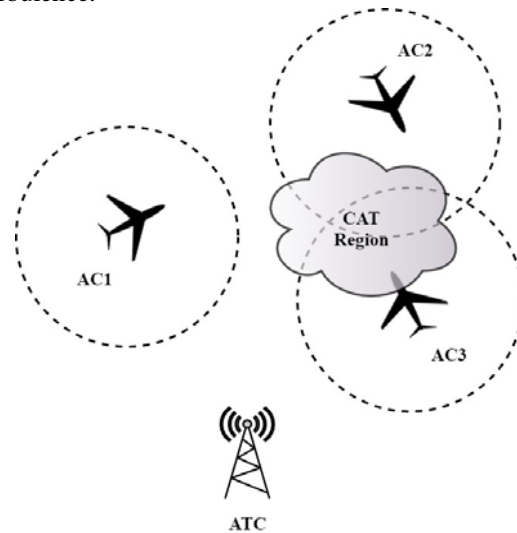

Fig. 1: Sample Flight Communication Scenario

## IV. CLEAR AIR TURBULENCE DETECTION SCENARIOS

In this Section we discuss the different scenarios that need to be considered while implementing our algorithms for avoiding clear air turbulence. CAT avoidance can be done either using communication via ATC tower or via direct communication between the IoT devices on aircraft.

In the first case of indirect communication, the ATC serves as the intermediary of information. Given the flight communication scenario as shown in Fig. 1, the time taken $T_I$, for a message from an aircraft that detects the CAT, denoted by $AC_{org}$ to reach target aircraft, denoted by $AC_{tar}$ is given by Equation 1, where $T_{ATC\_OH}$ denotes time overhead at ATC.

$$T_I = T_{AC\_org->ATC} + T_{ATC\_OH} + T_{ATC->AC\_tar} \quad (1)$$

Now, in the ideal scenario, the ATC has enough bandwidth to communicate with every aircraft at all times. In reality, this may not be the case, and ATC communicates with aircraft only at predefined intervals, or based on priority. In addition, the overhead at ATC consists of time to create a list of aircraft that are within the range and on possible path towards the CAT. Therefore, the total overhead possible at ATC, $T_{ATC\_OH}$, is given by Equation 2.

$$T_{ATC\_OH} = T_{ATC_{Interval}} + T_{ATC_{priority}} + T_{ATC_{List}} \quad (2)$$

In the second case, using the IoT framework for commercial aviation, there would be direct communication between aircraft sharing a region of the airspace. In this case, the first scenario is when the communication channel is always open for flights within communication range. It can be assumed that the IoT devices on flights can broadcast messages that can be received by other IoT devices within range. In the second scenario, the communication channel has to be established between flights within range to be able to exchange data. In the third scenario, the communication channel is established only on demand, and then the message is sent. Therefore, the time to send the alert directly $T_D$ between aircraft, is given by Equation 3.

$$T_D = T_{Channel_{Estd}} + T_{AC\_org->AC\_tar} \quad (3)$$

The time to establish the channel $T_{Channel\_Estd}$ is not applicable to the first scenario where communication channel is always open.

**Algorithm 1: Dectection of CAT**
**Input:** $AC_{Lat}, AC_{Long}, AC_{Alt}$
**Output:** CAT detection alert message
**begin**
    CheckSensorValues ($AC_{Lat}, AC_{Long}, AC_{Alt}$);
    $\delta_{sensor} \leftarrow |Sensor_{current} - Sensor_{avg.}|$;
    **if** $delta_{sensor} \geq Sensor_{Threshold}$ **then**
        Output ← SendMessage (Alert);
    **else**
        Output ← SendMessage (NULL);
    UpdateSensorValues ();
    Exit ();

## V. ALGORITHMS FOR CAT DETECTION AND AVOIDANCE

Whenever an aircraft experiences CAT, there is an option to notify the respective ATC using conventional technique over the radio to convey the CAT location to the ATC, which in turn would convey it to other aircraft in the same airspace. However, this conventional technique, would require manual intervention, and is much slower than an automated detection technique using the IoT framework.

The scenarios presented above consider communication between aircraft directly or via ATC. In all the cases, the assumption is there is a single ATC, which is in range of all the aircraft being currently considered. For a single ATC case, as shown in Fig 1, Algorithm 2 is used to send the CAT alert between aircraft via the ATC. However, in reality, this is not the case, and there exists multiple ATCs. Aircraft connect to the nearest ATC and exchange information with it. ATCs can connect to each other as well. During a flight, aircraft can connect to multiple ATCs one at a time, and handoffs happen when aircraft move from the range of one ATC to another.

**Algorithm 2: CAT alert via ATC**
**Input:** Flight path
**Output:** CAT detection alert message
**begin**
    **forall** $T_i \in Flight_{time}$ **do**
        Algorithm1 ($AC_{Lat}, AC_{Long}, AC_{Alt}$);
        **if** detectedCAT **then**
            **if** detectNoCommZone **then**
                storeData ();
                sendDataWhenCommEstd ();
            **else**
                **if** commOpenATC **then**
                    sendLoc ($AC_{Lat}, AC_{Long}, AC_{Alt}$);
                **else**
                    storeData ();
                    estdCommATC ();
                    sendLoc ($AC_{Lat}, AC_{Long}, AC_{Alt}$);
    receiveMsgATC ();
    createListATC ();
    **if** *BandwidthAvailable* **then**
        **forall** $AC_i \in List\{\}$ **do**
            sendLoc ($AC_{Lat}, AC_{Long}, AC_{Alt}$);
    **else if** *predefinedTimeInterval* **then**
        **forall** $AC_i \in List\{\}$ **do**
            sendLoc ($AC_{Lat}, AC_{Long}, AC_{Alt}$);
    **else if** *MessagePriorityATC* **then**
        **forall** $AC_i \in List\{\}$ **do**
            sendLoc ($AC_{Lat}, AC_{Long}, AC_{Alt}$);
    **else**
        waitATC ();
Exit ();

Fig. 2 depicts the multiple ATC scenario. There are 3 aircraft in this sample case, AC1, AC2 and AC3; in addition, there are 2 ATC towers ATC1 and ATC2. Out of the 3 aircraft shown, AC1 and AC2 are connected to ATC1 and AC3 is connected to ATC2. Also, the CAT region is depicted as well; at the given instant, AC3 has encountered CAT as detected using Algorithm 1; following the flight path, AC2 and AC3 would be entering the CAT region after certain time.

In this case, the data transfer would follow 4 steps. CAT detection data would be generated at AC3. Since there are no other aircraft within the region that is connected to ATC2, the message cannot be delivered directly from ATC2 to other aircraft. The data from AC3 would be transferred to communicating tower ATC2. The tower ATC2 would then forward the message to tower ATC1, which in turn would relay the message to AC1 and AC2. This scenario of multiple ATC towers in given in Algorithm 4.

As evident from the multiple ATC tower scenario given in Fig. 2, the alert message is sent over multiple hops and traverses longer distance as compared to a direct communication between AC3 and AC2 or between AC3 and AC1. In case of direct CAT

```
Algorithm 3: Direct CAT alert between aircraft
Input: Flight path
Output: CAT detection alert to aircraft within region
begin
    forall T_i ∈ Flight_time do
        createListNearbyAircrafts ();
        Algorithm1 (AC_Lat, AC_Long, AC_Alt);
        if detectedCAT then
            if CommChannelOpen then
                forall AC_i ∈ List{} do
                    sendLoc (AC_Lat, AC_Long, AC_Alt);
            else if CommChannelOnDemand then
                estdCommChannel ();
                forall AC_i ∈ List{} do
                    sendLoc (AC_Lat, AC_Long, AC_Alt);
            else if CommChannelEstablishedAlready then
                forall AC_i ∈ List{} do
                    sendLoc (AC_Lat, AC_Long, AC_Alt);
                checkNewAircraftRegion ();
                if AnyNewAircraft then
                    forall AC_i ∈ NewList{} do
                        sendLoc
                        (AC_Lat, AC_Long, AC_Alt);
            else
                storeDataDevice ();
    Exit ();
```

```
Algorithm 4: CAT alert in multiple ATC scenario
Input: Flight path
Output: CAT detection alert message
begin
    forall T_i ∈ Flight_time do
        Algorithm1 (AC_Lat, AC_Long, AC_Alt);
        if detectedCAT then
            if detectNoCommZone then
                storeData ();
                sendDataWhenCommEstd ();
            else
                findConnectedATC ();
                sendConnectedATCData ();
                exchangeATCData ();
                createListATC ();
                forall AC_i ∈ List{} ∈ CAT_Region do
                    sendLoc (AC_Lat, AC_Long, AC_Alt);
    Exit ();
```

alert between aircraft using the IoT framework is given in Algorithm 3.

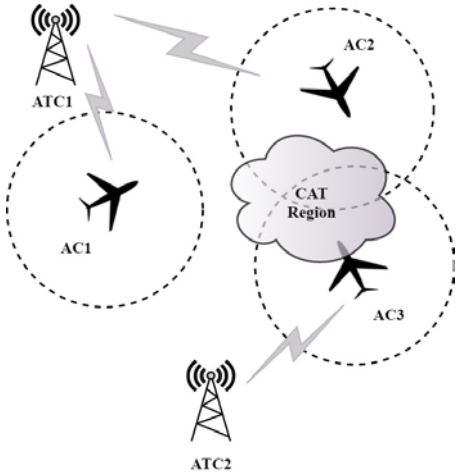

Fig. 2: Multiple ATC Towers Scenario

## VI. EXPERIMENTAL RESULTS

All the algorithms introduced in this paper are implemented and tested using simulation programs. The results for the scenarios of indirect and direct communication between aircraft are presented here. For all the scenarios, we measure the difference between the time of CAT detection and the time when the alert is received by the last aircraft within the region of airspace under consideration. This data provides a measure that can be used to compare the algorithms for the different scenarios. All the timings are reported in seconds.

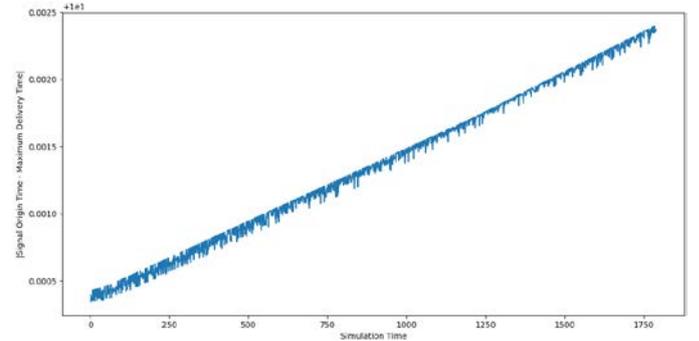

Fig. 3: Indirect Communication: Tower Broadcasts Signals

The implementation of the scenario of indirect communication, with the assumption of ATC being able to broadcast signal to every aircraft at all times is shown in Figure 3. As evident from Figure 3, the average difference between the signal's origin and maximum delivery time is increasing as the simulation goes on. This is to be expected as all aircraft in the simulation begin in same $1 \times 10^{10}\,m^2$ area but then begin to move in random directions as the simulation continues. Thus as the distance between the aircraft increases so does the average difference since the signal takes longer to propagate from the source to all other aircraft.

The scenario of indirect communication using the ATC where the signal from the ATC is broadcast at predefined interval is implemented as shown in Figure 4. The predefined interval for the simulation is chosen to be 50 seconds. In this case, it is evident that the average difference depends on when the signal originates. If the CAT is detected close to the end of the communication interval, the difference between the signal's origin time and its maximum delivery time is smaller as it must wait less time to be delivered by the tower.

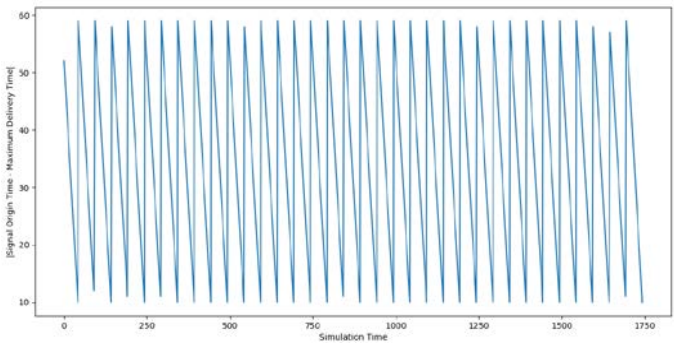

Fig. 4: Indirect Communication: Tower Broadcasts Signals in 50 second Intervals

The implementation of the third scenario for indirect communication, where the tower operates on data stored in a priority queue, is shown in Figure 5. It can be observed that the difference between a signal's origin and maximum delivery time is determined by the density of signals created. This is due to the priority queue at every tower getting longer as more signals are received. Thus the peaks in Figure 5 represent sections in time wherein a large portion of aircraft encountered turbulence.

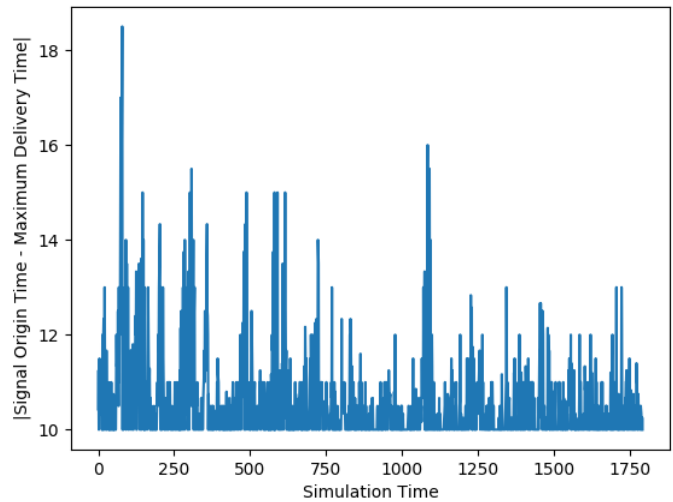

Fig. 5: Indirect Communication: Tower Broadcasts Signal at the Front of Priority Queue

The scenario of direct communication between aircraft using a broadcast of the CAT alert message is implemented and shown in Figure 6. From the graph it is evident that there is a linear relationship between simulation duration and max-origin difference times; this is due to the fact of increasing distance as a result of aircraft movement as the simulation progresses.

The results of the simulation for the scenario where the aircraft communicate using direct connection is shown in Figure 7. The graph is similar in terms of performance as compared to Figure 6 except for a small delay due to the signals being sent via a direct connection to each aircraft rather than a general broadcast.

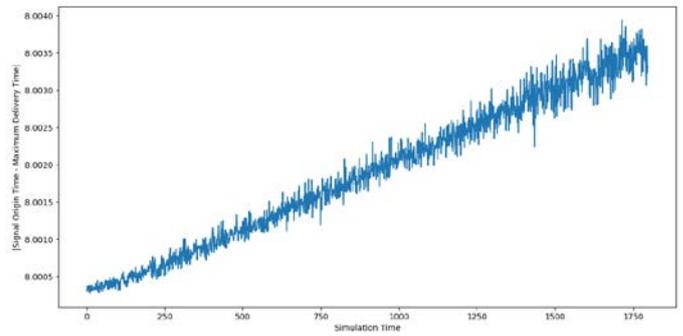

Fig. 6: Aircraft to Aircraft Communication: Signal Source Broadcasts Signal

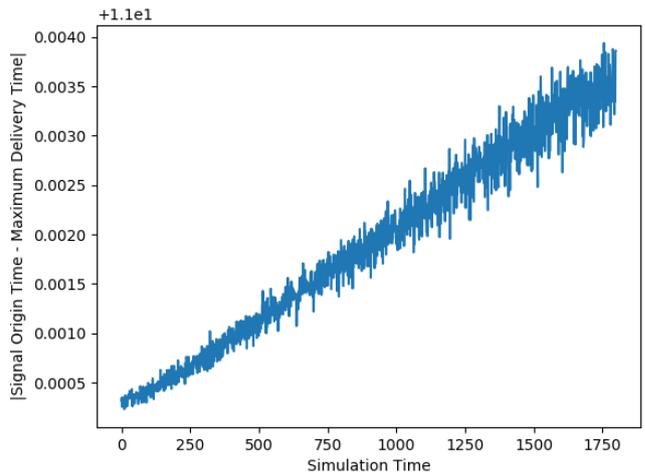

Fig. 7: Aircraft to Aircraft Communication: Signal Source Uses Open Connections to other Aircraft to Send Signals

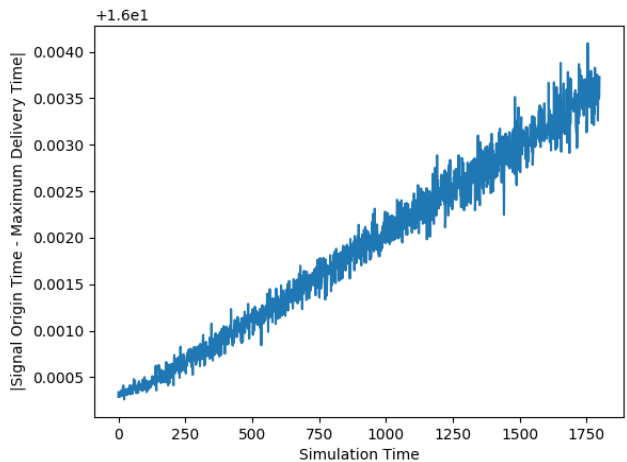

Fig. 8: Aircraft to Aircraft Communication: Signal Source creates Connections to other Aircraft to Send Signals

For the direct communication scenario between aircraft where establishing a connection is required, the implementation results are given in Figure 8. The main factor in this scenario is the time it takes to establish a new connection to an aircraft. As this time will be relatively constant for all aircraft, the primary factor affecting this graph will be the propagation delay of the signal. This is further evident in the scale of the figure with the fast average signal arriving just 0.0035 seconds ahead of the slowest signal.

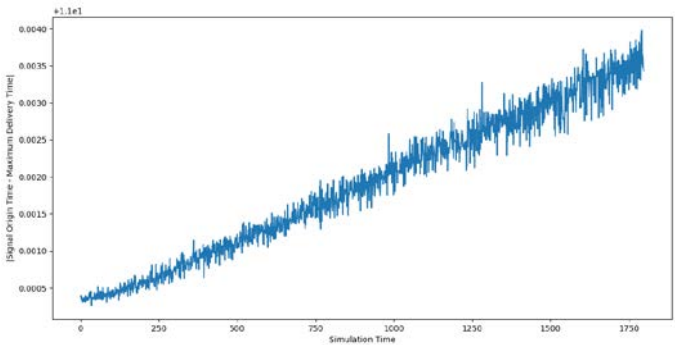

Fig. 9: Indirect Communication: Signal Sent via Multiple ATC Towers that are connected

The scenario of communication between aircraft in the presence of multiple ATC towers is implemented and the results are given in Figure 9. In terms of performance, these results are similar to those of the broadcast scenarios from the aircraft detecting the CAT.

Among the case of indirect communication, the scenario for broadcasting signal through the tower is the most efficient as turbulence data would be prioritized and broadcast to all other aircraft as soon as the data reaches a tower. Compared to the other scenarios for indirect communication however, this may not be the most feasible as it requires an open communication channel to be sustained between all aircraft and a nearby tower. As this is not possible for some international flights, the broadcast scenario might not be possible to implement universally despite it being the most efficient.

Among cases for direct communication, the scenario for broadcast from the aircraft detecting the CAT appears to be the most efficient as there is no significant delay between the time a signal is created and when it is broadcast to all other aircraft. While this makes the scenario time efficient, it does not guarantee the data reaches any other planes like the scenario using multiple ATC towers does.

Overall the scenario for direct communication with all aircraft within the region using broadcast message would be the most efficient in delivering signals to their targets in the least amount of time as there is no tower delay. Hence, using the IoT framework for detecting and avoiding CAT is better than techniques involving message delivery via ATC towers.

## VII. CONCLUSION

Safety of passengers, crew and the aircraft is the top priority in commercial aviation. Therefore, avoiding any type of turbulence during a flight is relevant. In this paper we introduce algorithms for clear air turbulence avoidance once it is detected by one aircraft. The introduced IoT model considers different scenarios, involving both direct and indirect communication using devices on the aircraft. From our experimental results, it can be concluded that direct communication between aircraft using IoT model is able to detect CAT more efficiently than any other scenario. Our future work would focus on other applications that can benefit from the IoT device communication between aircraft.